\documentclass[3p,times,procedia]{elsarticle}
\usepackage{nupha_ecrc}
\usepackage{graphicx,color}
\usepackage{amsmath}
\usepackage{amssymb}

\newcommand \beq{\begin{eqnarray}}
\newcommand \eeq{\end{eqnarray}} 
\def\x{{\boldsymbol x}}

\def\p{{\boldsymbol p}}

\newcommand{\rmd}{{\rm d}}
\newcommand{\rme}{{\rm e}}

\newcommand{\del}{\partial}

\volume{00}

\firstpage{1}

\journalname{Nuclear Physics A}

\runauth{J.-P. Blaizot et al.}

\jid{nupha}

\jnltitlelogo{Nuclear Physics A}

\usepackage{amssymb}
\usepackage[figuresright]{rotating}

\begin{document}

\begin{frontmatter}



\dochead{}
%% e.g. \dochead{17th International Conference on Dynamical Processes in Excited States of Solids}

\title{The subtle interplay of elastic and inelastic collisions in the thermalization of the quark-gluon plasma}


\author[cea]{Jean-Paul Blaizot}
\author[ind,rbrc]{Jinfeng Liao}
\author[wu]{Yacine Mehtar-Tani}

\address[cea]{Theoretical Physics, CEA, Saclay, France}
\address[ind]{Physics Department and Center for Exploration of Energy and Matter,
Indiana University, 2401 N Milo B. Sampson Lane, Bloomington, IN 47408, USA}
\address[rbrc]{RIKEN BNL Research Center, Bldg. 510A, Brookhaven National Laboratory,
   Upton, NY 11973, USA}
\address[wu]{Institute for Nuclear Theory, University of Washington, 
Seattle, WA 98195-1550, USA}

\begin{abstract}
Using kinetic theory, we analyze the interplay of elastic and inelastic collisions in the thermalization of the quark-gluon plasma.  The main focus is the dynamics and equilibration of long wavelength modes. 
\end{abstract}


\end{frontmatter}

% \linenumbers

\section{Introduction}

Understanding how the system of gluons that are created in the early stage of a heavy ion collision evolves into a locally equilibrated system remains an important and challenging problem, with many interesting facets and open issues (see Ref.~\cite{Berges:2012ks,Huang:2014iwa} for reviews -- a more up-to-date discussion is presented by A. Kurkela at this conference \cite{KurkelaQM}). Sticking to weak coupling approaches, much of the basic physics has been identified over the last decade or so. This involves the plasma instabilities, and their potential role in isotropizing the momentum distribution, the elastic and inelastic scatterings, and among the latter the importance of soft radiation \cite{Baier:2000sb,Xu:2004mz},  the role of the longitudinal expansion, etc. There exist detailed \cite{Baier:2000sb}, and very detailed \cite{Kurkela:2011ti},  parametric analysis at weak coupling  of these various physical processes. There are also many numerical  calculations, using either statistical classical field theory or kinetic equations. 

The present contribution reports on work based on kinetic theory.  We feel that there is room, aside from numerical calculations and parametric estimates,  for  understanding based on (differential) equations that we can control (semi) analytically. Our goal in these studies is  to identify robust, generic qualitative  behaviors that emerge from solutions of simple kinetic equations.  We are well aware of potential limitations of kinetic theory in the present context, especially in the description of the very longwavelength modes. However, the vast body of works on the thermalization of the quark-gluon plasma referring directly or indirectly to kinetic theory, as well as the  insight gained by simple analytical solutions,  justifies in our view the present exploration, and the postponement of its rigorous justification to a later stage.

\section{General setting}
We  describe the system of gluons\footnote{We ignore quarks in the present discussion.} that is produced in the initial stage of an ultra-relativistic heavy ion collision using kinetic theory with a phase-space distribution function  $f(\x,\p)$  which is independent of spin and color, uniform in space,  and  isotropic in momentum space\footnote{We assume that isotropization takes place over shorter time scales than those involved in the processes that we discuss in this paper.}. The  initial condition is inspired by the color glass (CGC)  picture \cite{Blaizot:2011xf}, $ f( p ) = f_0 \, \theta(1 - p/Q_s )$,
with $Q_s$ the saturation scale.
The final state is given by a Bose distribution of the form
$f(p)=[{\rme^{(p-\mu_{\rm eq})/T_{\rm eq}}-1}]^{-1}$,
with $T_{\rm eq}$ and $\mu_{\rm eq}$ respectively the equilibrium temperature and chemical potential. A non vanishing $\mu_{\rm eq}$ appears only when number changing processes can be neglected. Then $\mu_{\rm eq}$ can be either negative (underpopulation) or zero (overpopulation), with in the latter case formation of a condensate.  In all cases where $\mu_{\rm eq}=0$,  the equilibrium distribution is completely determined by the temperature, which is itself fixed by the initial condition, and energy conservation. 
 
 For the chosen initial condition,  the initial energy density $
 \epsilon_{\rm in} = {f_0 Q_s^4}/{(8\pi^2)}$ matches the equilibrium value, $\epsilon_{\rm eq}=T_{\rm eq}^4{\pi^2}/{30}$, for 
 $
 T_{\rm eq} = {f_0^{1/4} (15/4)^{1/4}} Q_s/{\pi} \approx 0.44 f_0^{1/4} Q_s.
$
 For realistic values of $f_0$, this temperature is less than $Q_s$. It is only for a very large overpopulation, i.e., 
 when $f_0\gtrsim 26$,  that $T_{\rm eq} \gtrsim Q_s$. 
The initial particle number is $
 n_{\rm in} = {f_0 Q_s^3}/({6\pi^2})$, while in equilibrium for $\mu_{\rm eq}=0$, $ n_{\rm eq} = {\zeta(3) T_{th}^3}/{\pi^2}  \propto f_0^{3/4}$. 
 The ratio 
 $ {n_{\rm in}}/{n_{\rm eq}} = [{f_0^{1/4} \pi^3}]/[{6 \zeta(3) (15/4)^{3/4}}] 
$
  is unity when $f_0 = f_c \approx 0.154$. Depending on whether $f_0>f_c$ or $f_0<f_c$,  the number density has to decrease or increase, respectively, in order for the system to reach equilibrium.

\section{Elastic scattering alone}

The approach to equilibrium with elastic scattering alone is described by the Boltzmann equation in the small angle approximation, i.e., a  Fokker-Planck equation of the form
\beq\label{FPeqn}
\frac{\del f(\tau,p)}{\del \tau}=-\frac{1}{p^2}\frac{\del}{\del p}\left( p^2 {\mathcal J}(\tau,p)  \right),\qquad 
{\mathcal J}=-  \left[ I_a\, \partial_p f + I_b\, f(1+f)\right] . \eeq
In the expression of the current ${\mathcal J}(\tau,p)$,  $I_a$ and $I_b$ are the following integrals
\beq\label{definitionsI}
I_a\equiv \int\frac{d^3 p}{(2\pi)^3} f(p)(1+f(p)),\qquad I_b\equiv \int\frac{d^3 p}{(2\pi)^3}\frac{2f(p)}{p}.
\eeq 
The time $\tau$ is related to the physical time $t$ by 
$
\tau\equiv 4 \pi \alpha^2N_c^2  t{\cal L}Q_s=4\pi^3 \bar\alpha^2 \;t {\cal L}Q_S,
$
where ${\mathcal L}=\int \rmd q/q$ is the Coulomb logarithm, treated here as a constant, and all momenta are expressed in units of $Q_s$. 
   
\begin{figure}[!hbt]
\begin{center}
\includegraphics[width=0.35\textwidth]{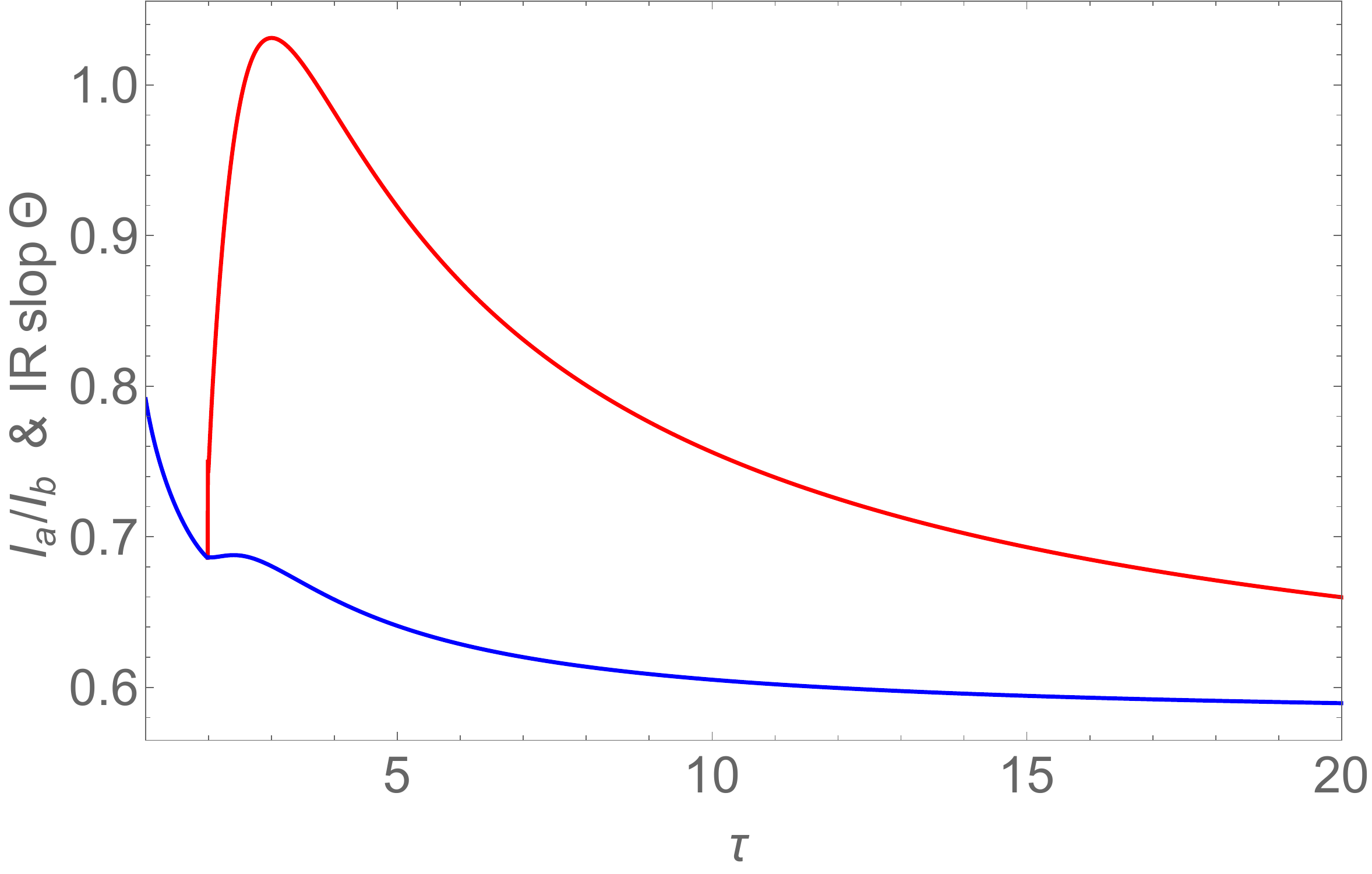}
% \vspace{-0.1in}
\caption{(Color online). The (blue) lower curve represents the effective temperature $T^*=I_a/I_b$ as a function of time. The (red) upper curve represents the coefficient $\Theta$. Before onset,  the distribution function is regular at the origin and $\Theta=0$. At onset,  $\Theta=T^*$. After the onset of condensation, a flow of particles quickly develops at the origin of momentum space and the system is rapidly driven away from local equilibrium, with $\Theta$ becoming significantly larger than the local effective temperature $T^*=I_a/I_b$ which stays almost constant across the transition. At later times, $\Theta$ decreases and eventually converges back to $T^*$.  The calculation is done for moderate overpopulation, $f_0=1$. As $f_0$ increases more and more particles tend to condense, and as a result the growth of the coefficient $\Theta$ just beyond onset is more important.  } \vspace{-0.25in}
\label{fig:localglobal}
\end{center}
\end{figure}

Thermalization proceeds then with conserved particle number.  A chemical potential quickly develops, and, at small momenta, the solution acquires the form of an equilibrium distribution, $f(p)\simeq T^*/(p-\mu^*)$, with $T^*$ and $\mu^*$ time dependent parameters, with $T^*=I_a/I_b$ and $\mu^*<0$. As time passes, the chemical potential increases and eventually vanishes at the onset of Bose condensation \cite{Blaizot:2013lga}. Beyond the onset\footnote{Equations that govern the evolution of the condensate in the small angle approximation \cite{Blaizot:2015xga}, including finite mass effects\cite{Blaizot:2015wga}, have recently been obtained. However,  the simple procedure that we follow here is sufficient for the present discussion.} there exists a solution\cite{Blaizot:2014jna} that behaves at small $p$ as $\Theta/p$, which allows for a flux $ {\cal F}$ of particles at the origin,  $
  {\cal F}(p=0)=(4\pi I_a/T^*) \left( T^*-\Theta\right)\Theta.
 $
At the onset of condensation,  $\Theta=T^*$. Just after onset, $\Theta$ becomes rapidly larger than $T^*$, producing a negative flux of particles at the origin. This is illustrated in Fig.~\ref{fig:localglobal}.   
  
      % \vspace{-0.05in}
  
   The non-local character of the kinetic equation Eq.~(\ref{FPeqn})  is worth-emphasizing: although it looks like a local partial differential equation for the function $f(\tau,\p)$, there is in fact a non-linear coupling with the entire solution through the integrals $ I_a$ and $ I_b$. These integrals encode in particular the information of whether the system is under or over populated.  Just beyond the  onset of condensation, however, $I_a$ and $I_b$ vary only mildly, while the system is rapidly driven out of equilibrium, a behavior largely determined then by local properties. 
  
\section{Inelastic scattering alone}

\begin{figure}[!hbt]
\begin{center}
\includegraphics[width=0.35\textwidth]{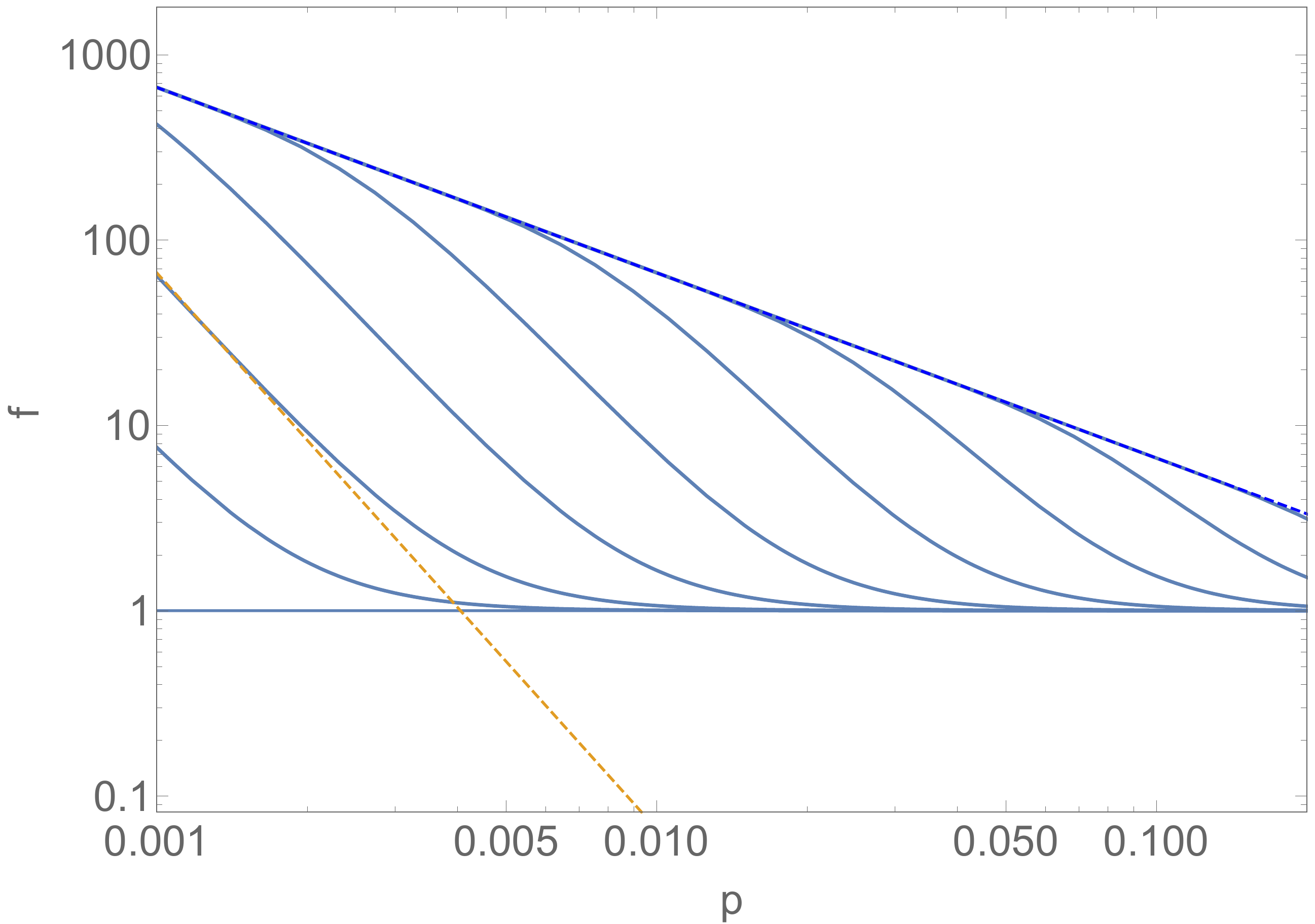}
 \vspace{-0.1in}
\caption{(Color online). The solution $f(p,\tau)$ of Eq.~(\ref{solutioninel}) at small momenta and short times. The fixed point solution $T^*/p$ is indicated by the blue dashed line. The wave front of the diffusion wave (centered at $P_*\sim \sqrt{\tau}$) has the shape of a transient radiation spectrum $\sim 1/p^{-3}$ (yellow dashed line). The horizontal line indicates the initial condition $f_0=1$.} \vspace{-0.25in}
\label{fig_1}
\end{center}
\end{figure}

Thermalization can also be achieved by inelastic processes alone, which are dominated by the emission or the absorption of soft gluons. To treat those processes, we exploit recent progress in their analytical understanding in our work on medium induced cascades \cite{Blaizot:2015jea}.
We describe those with a  generalization of an equation that has been studied in \cite{Blaizot:2015jea}, that uses a simplified splitting kernel, and that can be brought to the form\footnote{It is also a simplified version of the equation used in Ref.~\cite{Kurkela:2014tea}
.}
\beq
 \label{eq_D_BH_5}
\partial_\tau f(p,\tau)= \frac{RT^*}{ p^3}  \left\{ \int_0^\infty \rmd k K(p,p+k) \Phi(p,p+k) -\int_0^p\rmd k K(k,p) \Phi(k,p)\right\},\qquad K(p,p')=\frac{p'^3}{p'-p},
\eeq
where 
 $
\Phi(p,p')\equiv f(p')+f(p')f(p)+f(p'-p)\left[ f(p')-f(p)  \right].$
The quantity $R$ is a parameter, generically of order unity \cite{Huang:2013lia}, that we keep as a free parameter to control the relative strength of elastic and inelastic scatterings. 
A simple analysis reveals that in the small $p$ region, this equation reduces approximately to
\beq
\partial_\tau f(p) \approx\frac{RI_a}{p^3}\left[ T^*-pf(p)   \right],
\eeq 
whose solution  is easily obtained. When one ignores the time dependence of $I_a$ and $T^*$, it reads
\beq\label{solutioninel}
f(p,\tau)= \frac{T^*}{p}+\left( f_0- \frac{T^*}{p} \right)\rme^{-P_*^2(\tau)/p^2} ,\qquad P_*^2(\tau)=RI_a \tau,\qquad T^*=\frac{I_a}{I_b}.
\eeq
 One sees that the  small momentum behavior is controlled by an essential singularity, which implies that the fixed point $\sim 1/p$ is reached in no time. As time progresses the fixed point solution is populated via a diffusion wave centered at $P_*(\tau)$, as illustrated in Fig.\ref{fig_1}.
 
% \vspace{-0.05in}
 
 This is in complete consistency from what one obtains from exactly solving the full kernel numerically.

%%%%%%%%%%%%%%%%%%%%%%%%%%%%
\section{Interplay of elastic and inelastic scattering}
%%%%%%%%%%%%%%%%%%%%%%%%%%%%%

 Common to both types of processes is the rapid growth of the low momentum modes, and in the case of overpopulated systems, the disappearance of particles at low momenta where the distribution function behaves approximately as $\sim 1/p$. However the way this is achieved  is different in the elastic and the inelastic cases. In the elastic case, the $1/p$ behavior develops gradually, via the building of a strong current of particles towards small momenta, leading eventually to Bose condensation that eliminates the particles that cannot be accomodated in the thermal spectrum. In the inelastic cases, the population of soft modes is driven by radiation, which leads to a rapidly growing spectrum $\sim 1/p^3$, forcing local equilibrium to be reached instantly at small $p$. 

When both processes are present, it appears that the small $p$ dynamics remain dominated by the inelastic ones. The fast radiation of soft gluons, and the accompanying absorption,  force the distribution to behave as $1/p$ in no time. One might think that this could facilitate Bose condensation \cite{Huang:2013lia}, as such a behavior is a prerequisite. However, the very reasons that the inelastic processes force local equilibrium near $p=0$ imply a fast suppression of the excess particles, thereby hindering condensation,  making it in fact unnecessary. What is involved here is not so much the fact that the inelastic processes may be globally of comparable magnitude as the inelastic ones (i.e. $R\approx 1$). The major factor is radiation, which allows transport of momenta over a wide range of  momenta and  on a very short time scale. This is a specific feature of gauge theories, that do not show up in scalar theories for instance. 

The present note focussed on what happens near $p=0$, and how the  particles in excess  in the initial distribution are eliminated. The detailed pattern of the evolution towards equilibrium depends on the value of $R$. For instance if $R$ is small,  the bulk of the dynamics remains the same as in the absence of inelastic processes, the influence of inelastic processes being significant only in the small region $p<P_\ast$.  The  description of what happens for different values of $R$ will be presented in a forthcoming publicaiton.

 \section*{Acknowledgements}
 The research of JPB is supported by the European Research Council under the Advanced Investigator Grant ERC-AD-267258. The research of JL is partly supported by the NSF (Grant No. PHY-1352368) and by the RIKEN BNL Research Center



%% References


\bibliographystyle{elsarticle-num}
\bibliography{<your-bib-database>}






% \end{thebibliography}

\end{document}